\documentclass[preprintnumbers,amsmath,amssymb]{revtex4-1}

\usepackage{epsfig}
\usepackage{graphicx}
\usepackage{subfigure}
\usepackage{float}
\usepackage{multirow}
\usepackage{amssymb}
\usepackage{amsmath}
\usepackage{picinpar}

\begin{document}

\newcommand{\figwidth}{0.8\columnwidth}
\newcommand{\subfigwidth}{0.6\columnwidth}

\newcommand{\smin}{{\mbox{\scriptsize{min}}}}
\newcommand{\smax}{{\mbox{\scriptsize{max}}}}
\newcommand{\stot}{{\mbox{\scriptsize{tot}}}}
\newcommand{\tot}{{\mbox{\scriptsize{tot}}}}

\newcommand{\ms[2]}{{\mbox{\scriptsize{#2}}}}
\newcommand{\im}{i}
\newcommand{\jm}{j}

%%% System specific
\newcommand{\mA}{\langle m_A\rangle}
\newcommand{\mB}{{\langle m_B\rangle}}
\newcommand{\pmB}{{\langle pm_B\rangle}}
\newcommand{\m}{{\langle m_\tot\rangle}}

\title{Monte Carlo simulations of an Ising bilayer with non-equivalent planes}

\author{I. J. L. Diaz}\email{ianlopezdiaz@gmail.com}
\affiliation{Departamento de F\'{i}sica,
Universidade Federal de Santa Catarina,
88040-900, Florian\'{o}polis, SC, Brazil}
\author{N. S. Branco}\email{nsbranco@fisica.ufsc.br}
\affiliation{Departamento de F\'{i}sica,
Universidade Federal de Santa Catarina,
88040-900, Florian\'{o}polis, SC, Brazil}

\date{\today}

\begin{abstract}
We study the thermodynamic and magnetic properties of an Ising bilayer ferrimagnet.
The system is composed of two interacting non-equivalent planes
in which the intralayer couplings are ferromagnetic while the interlayer interactions are antiferromagnetic.
Moreover, one of the planes is randomly diluted.
The study is carried out within a Monte Carlo approach employing
the multiple histogram reweighting method and finite-size scaling tools.
The occurrence of a compensation phenomenon is verified and the compensation temperature,
as well as the critical temperature for the model,
are obtained as functions of the Hamiltonian parameters.
We present a detailed discussion of the regions of the parameter space where
the compensation effect is present or absent.
Our results are then compared to a mean-field-like approximation applied to the same model
by Balcerzak and Sza{\l}owski (2014).
Although the Monte Carlo and mean-field results agree qualitatively,
our quantitative results are significantly different.
\end{abstract}


\maketitle

\section{Introduction}\label{introduction}

Layered magnetic materials have attracted significant attention in the last few decades \cite{de1990magnetic}.
In particular, phenomena such as the giant magnetorresistance \cite{camley1989theory}
and the magnetocaloric effect \cite{phan2007review},
present in many of these materials, have important technological applications \cite{camley1989theory, phan2007review, levy1990electrical}.
There is also great theoretical interest in studying the magnetic properties of these systems,
in order to gain insight into the crossover between the characteristic behavior of two- and three-dimensional magnets. %\cite{araujo2002study}.

One particularly interesting issue in these magnetic systems is the
possible presence of a compensation temperature, in which the total magnetization is zero at a temperature below the critical one.
Moreover, the role of dilution (or disorder, in general) may be an important one, since pure systems are not ubiquitous in nature,
and the controlled growth of non-homogeneous materials may allow for the selection of desired physical behavior.
Therefore, we study the magnetic properties of an Ising system composed of two non-equivalent atomic layers,
each modeled as square lattices with ferromagnetic intralayer couplings.  
One of the layers is additionally randomly diluted and the coupling between both layers is antiferromagnetic.
Our main goal is to establish the necessary conditions for the presence or absence of the compensation temperature
and how it depends on the parameters of the Hamiltonian.
%Experimental \cite{smits2004antiferromagnetic, leiner2010observation, chung2011investigation}

From the theoretical point of view, exact solutions for magnetic models do exist,
but they are limited to %certain one- and two-dimension
a handful of cases \cite{baxter1982exactly}.
Therefore, the analysis of more complex models, with the introduction of different ferromagnetic and antiferromagnetic interactions,
as well as the presence of atomic disorder, requires approximate approaches.
Methods such as transfer matrix \cite{lipowski1993layered, lipowski1998critical},
renormalization group \cite{hansen1993two, li2001critical, mirza2003phenomenological},
mean-field (MF) approximations, and Monte Carlo (MC) simulations \cite{hansen1993two}
have been applied to an Ising bilayer without disorder.
A more sophisticated version of a mean-field approach, called pair approximation method (PA),
has also been used to study both bilayers and multilayers with Ising and
Heisenberg spins without disorder \cite{szalowski2012critical, szalowski2013influence},
as well as an Ising and Heisenberg bilayer with disorder \cite{balcerzak2014ferrimagnetism}.
The PA takes nearest-neighbor correlations into account
as opposed to a straightforward MF approach where all correlations are completely neglected.
However, the PA still neglects all correlations beyond nearest-neighbors' and,
despite giving a more precise estimate of the critical point than a standard MF approach,
the PA will overestimate the true critical temperature of the model.
Moreover, mean-field-like approximations may not be an adequate tool to describe the actual physical behavior of some systems
(see Ref. \onlinecite{boechat2002renormalization} and references therein).
Therefore, the use of more precise methods is necessary in order to establish the correct physical picture.

Since, to the best of our knowledge, no Monte Carlo calculations have been made for
either bilayer or multilayer models with disorder, in this work we present an analysis of
the thermodynamic and magnetic properties of disordered bilayers.
This analysis is carried out within a Monte Carlo approach, using Metropolis \cite{artigo:metropolis} and Wolff \cite{artigo:wolff} algorithms
and with the aid of a reweighting multiple histogram technique \cite{artigo:ferrenberg:histograma1, artigo:ferrenberg:histograma2}.
In Sec. \ref{model} we present and discuss the model for the magnetic bilayer.
The simulation and data analysis methods are discussed in Secs. \ref{monte_carlo} and \ref{data_analysis} respectively.
We present our results and discussion in Sec. \ref{results}.
In the last section, \ref{conclusion}, the final remarks and conclusions are drawn.

\section{Model}\label{model}

\begin{figure}[h]
\begin{center}
\includegraphics[width=\subfigwidth]{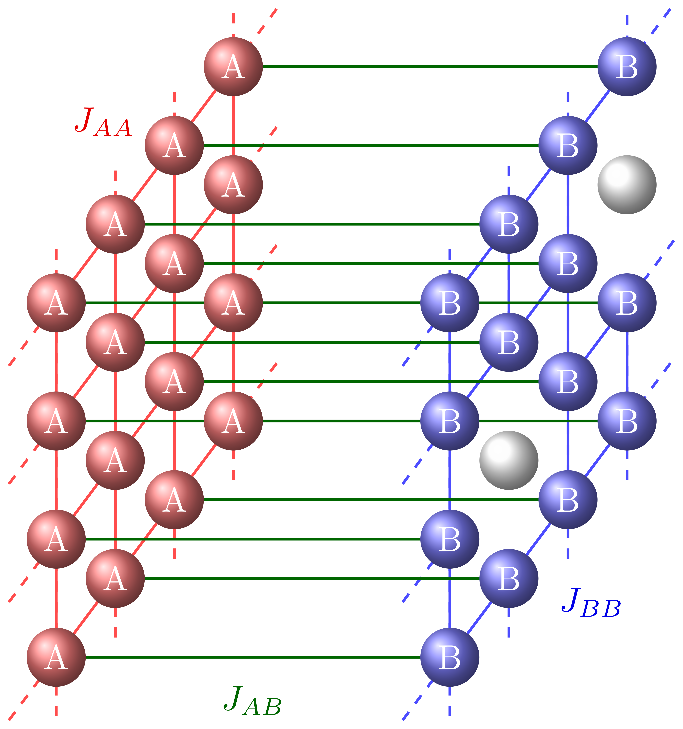}
\caption{
\label{fig_bilayer}
Schematic representation of the system composed by two layers, \textbf{A} and \textbf{B}.
The exchange integral between two neighboring atoms belonging to the same layer
is $J_{AA}>0$ for layer \textbf{A} and $J_{BB}>0$ for layer \textbf{B}.
The exchange integral between atoms belonging to different planes is $J_{AB}<0$.
Only layer \textbf{B} is randomly diluted.
}
\end{center}
\end{figure}

The system is composed of two monoatomic layers, \textbf{A} and \textbf{B} (see Fig. \ref{fig_bilayer}).
Both layers are modeled as square lattices of linear size $L$ with nearest-neighbor interactions and periodic boundary conditions.
The interaction between two neighboring atoms belonging to the same layer
is ferromagnetic whereas the interaction between atoms belonging to different planes is antifferromagnetic.
Layer \textbf{B} is randomly diluted to model the presence of a non-magnetic quenched impurity among the atoms of type \textbf{B}.

Each atom has a magnetic degree of freedom that is assumed to behave as an Ising-like spin,
so the system is described by a spin-1/2 Ising Hamiltonian as follows
\begin{align}\label{eq:hamiltonian}
\mathcal{H} =
%&
-\sum_{\langle i\in A,j\in A\rangle}J_{AA}s_i s_j
-\sum_{\langle i\in A,j\in B\rangle}J_{AB}s_i s_j\epsilon_j
%\nonumber \\
%&
-\sum_{\langle i\in B,j\in B\rangle}J_{BB}s_i s_j\epsilon_i\epsilon_j,
\end{align}
where the sums go over all nearest-neighbor pairs, the spin variables assume the values $s_i=\pm 1$ for all sites $i$,
and the $\epsilon$'s are quenched, uncorrelated random variables, chosen to be $1$ with probability $p$ (active site concentration),
or $0$ with probability $1-p$ (impurity concentration, or spin dilution).
The exchange integrals $J_{AA}$ and $J_{BB}$ are positive and $J_{AB}$ is negative.

%%
%% FIG 01: bilayer
%%

When performing the Monte Carlo simulations we calculate some observables
such as the dimensionless extensive energy
\begin{align}\label{eq:E}
E =
-\sum_{\langle i\in A,j\in A\rangle}(J_{AA}/J_{BB})s_i s_j
-\sum_{\langle i\in A,j\in B\rangle}(J_{AB}/J_{BB})s_i s_j\epsilon_j
-\sum_{\langle i\in B,j\in B\rangle}s_i s_j\epsilon_i\epsilon_j,
\end{align}
the magnetizations in planes \textbf{A} and \textbf{B}, respectively
\begin{align}
	\label{eq:mA}
	m_A=\frac{1}{N_A}\sum_{i\in A}s_i,
	\\
	\label{eq:mB}
	m_B=\frac{1}{N_B}\sum_{j\in B}s_j\epsilon_j,
\end{align}
and the total magnetization
\begin{equation}\label{eq:m}
	m_\tot
	%=\frac{1}{N}\left(\sum_{i\in A}s_i+\sum_{j\in B}s_j\epsilon_j\right)
	=\frac{1}{2}\left(m_A+pm_B\right),
\end{equation}
where
%$N=2L^2$ is the total number of sites on the system and
$N_A=L^2$ is the number of sites in plane \textbf{A} and $N_B=pL^2$ is the number of active sites in plane \textbf{B}.

With the above quantities, we are able to measure observables
of the form $\mathcal{O}=m_\Lambda^kE^\ell$,
where $k,\ell=0,1,2,3,\ldots$, and $\Lambda=A,B,\tot$.
We denote $\langle\mathcal{O}\rangle$ as the thermal average
for a single disorder configuration whereas we shall denote the subsequent average over disorder configurations
of $\langle\mathcal{O}\rangle$ as $\overline{\langle\mathcal{O}\rangle}$.
We also define the %specific heat
%\begin{equation}
%	c=\frac{K^2}{N_\tot}\overline{\left(\langle E^2\rangle-\langle E\rangle^2\right)},
%\end{equation}
%and the
magnetic susceptibilities
\begin{align}
	\label{eq:sus}
	\chi_\Lambda
	=N_\Lambda K\overline{\left(\langle m_\Lambda^2\rangle-\langle |m_\Lambda|\rangle^2\right)},
\end{align}
where $K=J_{BB}/(k_BT)$ is the inverse dimensionless temperature, and $\Lambda=A, B, \tot$.
%The number of sites in plane \textbf{A} and the number of active sites in plane \textbf{B} are, respectively, $N_A$ and $N_B$.
The total number of active sites in the system is $N_{\tot}=N_A+N_B$.

The quantities defined above were calculated by means of Monte Carlo simulations
and used to estimate the critical and compensation temperatures for the model,
as presented in Sec. \ref{monte_carlo} and Sec. \ref{data_analysis}.

\section{Simulational Details}\label{monte_carlo}

We studied the Ising bilayer described by Hamiltonian (\ref{eq:hamiltonian}) within a Monte Carlo approach.
We have employed both Metropolis \cite{artigo:metropolis} and Wolff \cite{artigo:wolff} algorithms
to simulate two interacting square lattices with $L^2$ sites each and periodic boundary conditions.
All sites on lattice \textbf{A} are active, while each site on lattice \textbf{B} is randomly chosen
to be active ($\epsilon_i=1$) or a vacancy ($\epsilon_i=0$) with probabilities $p$ or $1-p$, respectively.
The Metropolis dynamics was used for some simulations away from the critical point and the Wolff dynamics
was used for temperatures close to $T_c$ where it is more efficient.
All random numbers were generated using the Mersenne Twister pseudo-random number generator \cite{artigo:mersenne-twister}.

A Monte Carlo step per spin (MCS) corresponds to $2L^2$ spin updates for the Metropolis dynamics,
or a cluster flip for the Wolff dynamics.
Our simulations ran typically from $2\times 10^4$ to $5\times 10^7$ MCSs
and we made sure to generate at least $n=1000$ uncorrelated states,
with $n$ given by
\begin{equation}
	n=\frac{n_{\ms{MCS}}-t_{\ms{eq}}}{2\tau},
\end{equation}
where $n_{\ms{MCS}}$ is the number of MCSs, $t_{\ms{eq}}$ is the equilibration time and
$\tau$ is the largest correlation time.
For the Metropolis dynamics we use the integrated correlation time, $\tau_{\scriptsize{\mbox{int}}}$,
whereas for the Wolff dynamics all time scales have to be adjusted such that the real correlation time is given by
\begin{equation}
	\tau=\tau_{\scriptsize{\mbox{int}}}\times \frac{\langle n_c\rangle}{N_\tot},
\end{equation}
where $\langle n_c\rangle$ is the average cluster size \cite{livro:barkema}.

For each set of parameters ($J_{AA}/J_{BB}$, $J_{AB}/J_{BB}$, $p$, $L$),
we chose a range of temperatures of interest and divide this range in 5 to 15 temperatures to run simulations at.
We then use the multiple-histogram method \cite{artigo:ferrenberg:histograma1,artigo:ferrenberg:histograma2, livro:barkema}
to compute our observables $\mathcal{O}$ at any temperature inside this range.
The thermal error associated with those observables is estimated via the blocking method \cite{livro:barkema},
in which we divide the data from each simulation in blocks and repeat the multiple-histogram procedure
for each block. The errors are the standard deviation of the values obtained
for a given observable for different blocks. %%$\langle\mathcal{O}\rangle$
For each temperature we repeat the process for $N_s$ samples of quenched disorder to obtain
the final estimate of our observable, $\overline{\langle\mathcal{O}\rangle}$.
We chose $10\leq N_s\leq 50$, such that the error due to disorder was approximately
the same as the thermal error obtained for each disorder configuration.
Finally, we sum both thermal and disorder errors for an estimate of the total error.

\section{Data analysis}\label{data_analysis}

The compensation point is found by locating the temperature, $T_{comp}$, where the total magnetization is zero and
the plane magnetizations, $\mA$ and $\mB$, remain non-zero, as seen in Figs. \ref{fig:magLmax} and \ref{fig:mags}.
To estimate that temperature, for each disorder configuration, we perform simulations for a number of temperatures
around the $\m=0$ point and obtain the $\m$ values as a continuous function of $T$ using the multiple-histogram method.
We then find the root of $\langle m_\tot(T)\rangle$ using Brent's method \cite{artigo:brent1973}.
We repeat this for $N_s$ configurations to estimate the disorder error, similar to what is discussed in Sec. \ref{monte_carlo}.

It is important to point out that different values of $L$ give very close estimates of $T_{comp}$ as seen in Fig. \ref{fig:mags:a}.
The different $T_{comp}$ estimates oscillate with no discernible tendency as $L$ increases,
therefore, we average our results for different values of $L$, in addition to averaging over samples.
To estimate the error bars, we sum the standard deviations obtained over the $N_s$ samples, over different values of $L$,
and the error estimated for a single sample via the blocking method.
In some cases, particularly for $T_{comp}$ close to $T_c$, the small lattices give $T_{comp}$ estimates
that differ more than one error bar from the mean value.
In those cases, the smaller lattices are excluded from the analysis.

%%
%% FIG02a: magnetization
%% FIG02b: magnetization
%%

%%
%% FIG03a: magnetization
%% FIG03b: magnetization
%%

For an accurate determination of the critical point,
since in MC simulations we necessarily deal with finite systems
and the critical phenomena happen in the thermodynamic limit,
it is necessary to examine the size dependence of the observables
measured for finite systems of various sizes and extrapolate these results to the $L\rightarrow\infty$ limit.
In this finite-size scaling approach \cite{livro:julia} we write the singular part of the free energy density for a
system of linear size $L$ near the critical point as
\begin{equation}\label{eq:RGf2}
	\bar f_{\ms{sing}}(t,h,L)
	%\sim L^{-d}f^0(tL^{y_t},hL^{y_h},\{ \bar u_iL^{-\omega_i}\})
	%\sim L^{-d}f^0(tL^{1/\nu}, hL^{(\gamma+\beta)/\nu})
	\sim L^{-(2-\alpha)/\nu}f^0(tL^{1/\nu}, hL^{(\gamma+\beta)/\nu})
\end{equation}
where $t$ is the reduced temperature, $t=(T-T_c)/T_c$, $T_c$ is the critical temperature of the infinite system,
$h=H/(k_BT)$ and $H$ is the external magnetic field. The critical exponents $\alpha$, $\gamma$, $\beta$, and $\nu$
are the traditional ones associated with the magnetic susceptibility, magnetization and correlation length, respectively.

%Taking appropriate derivatives of the free energy, it is possible to show that
%some thermodynamic quantities exhibit similar scaling forms at $h=0$:
%\begin{align}
%	m_\tot &=L^{-\beta/\nu}\mathcal{M}(x_t),
%	\label{FSS_mag}
%\\
%	\chi_\tot &= L^{\gamma/\nu}
%	\mathcal{X}(x_t),
%	\label{FSS_chi}
%\\
%	c &= L^{\alpha/\nu}
%	\mathcal{C}(x_t),
%	\label{FSS_c}
%\end{align}
%where $x_t=tL^{1/\nu}$ is the temperature scaling variable.

%Some quantities, such as $\chi_\tot$ or $c$ (only if $\alpha>0$), diverge at the critical point as $L\rightarrow\infty$,
%as it is clear from the scaling law in Eqs. (\ref{FSS_chi}) and (\ref{FSS_c}).
%For a finite system, however, each of these quantities has a peak at a temperature, $T_c(L)$,
%which is the pseudo-critical transition temperature.

Taking appropriate derivatives of the free energy, it is possible to show that
some thermodynamic quantities exhibit similar scaling forms at $h=0$.
Some of these quantities diverge at the critical point as $L\rightarrow\infty$,
as it is the case for the magnetic suceptibility, the specific heat (only if $\alpha>0$),
the thermal derivative of the Binder cumulant or other thermal derivatives \cite{artigo:landau}.
For a finite system, however, each of these quantities has a peak at a temperature, $T_c(L)$,
which is the pseudo-critical transition temperature.
For the magnetic susceptibility we have the following scaling form
\begin{equation}
	\chi_\tot = L^{\gamma/\nu}
	\mathcal{X}(x_t),
	\label{FSS_chi}
\end{equation}
where $x_t=tL^{1/\nu}$ is the temperature scaling variable.
As it is clear from Eq. (\ref{FSS_chi}),
$\chi_\tot$ diverges at the critical point as $L\rightarrow\infty$
and, for a finite system size, the maximum occurs when
\begin{equation}
	\left.\frac{d\mathcal{X}(x_t)}{dx_t}\right|_{T=T_c(L)}=0,
\end{equation}
which gives us the following scaling law
\begin{equation}
	\label{eq:FSS:tc}
	T_c(L) = T_c+AL^{-1/\nu},
\end{equation}
where $A$ is a constant, $T_c$ is the critical temperature and $\nu$ is the
critical exponent associated with the correlation length.

It is important to point out that the finite-size scaling method
based on the peaks of different quantities is expected to give consistent results,
as we were able to verify in preliminary simulations.
In this work, however, we focused only on the peak temperatures of the
magnetic susceptibilities, defined in Eq. (\ref{eq:sus}),
for these peak temperatures occurred fairly close to one another and were the
sharpest peaks from all the quantities initially considered.

For the location of the peak temperature we use the multiple-histogram method.
This procedure is also automated and the maximum is found using the Broyden-Fletcher-Goldfarb-Shanno (BFGS) method \cite{artigo:BFGS}.
Fig. \ref{fig:mhist} shows an example of the use of the multiple-histogram method to obtain the
magnetic susceptibility $\chi_\tot$ as a continuous function of temperature, which enables us to locate the peak temperature for that particular observable.

After we have the estimates for $T_c(L)$, we fit the data to Eq. (\ref{eq:FSS:tc}).
This equation has three free parameters to be adjusted in the fitting process and requires great statistical resolution
in order to produce stable and reliable estimates of the parameters.
It is also possible to obtain an independent estimate of $\nu$ via other finite-size scaling laws and use this value
in Eq. (\ref{eq:FSS:tc}), effectively reducing the number of free parameters by one.
However, this approach has a downside as the \emph{maxima} of more thermodynamic quantities would have to be evaluated
and these \emph{maxima} do not always happen to be close to one another.
Moreover, it is usually necessary to simulate really large systems in order to obtain a reliable estimate of critical exponents. 
This means we would need more simulations for a wider range of temperatures and, consequently, more computational work.

%%
%% FIG04: histogram method
%%

%%
%% FIG05: critical temperature
%%

Since we are more interested in obtaining the critical temperature than in finding a precise value for the exponent $\nu$,
we employ a procedure similar to the one presented in Refs. \onlinecite{artigo:landau} e \onlinecite{diaz2012feru},
in which we set a fixed value for the exponent $\nu$ and perform fits with two free parameters, instead of three.
These fits are made, for a fixed value of $\nu$, for system sizes not smaller than $L_{\smin}$
and the value of $L_\smin$ that gives the best fit is located, i. e.,
the one that minimizes the reduced weighted sum of errors $\chi^2/n_{DOF}$,
where $n_{DOF}$ is the number of degrees of freedom.
Next, we keep changing the values of $\nu$ and $L_\smin$ iteratively until we locate the set of values
that globally minimizes $\chi^2/n_{DOF}$. Examples of these fits are shown in Fig. \ref{fig:tc}.
This procedure effectively linearizes the fit, although it does not allow for
an individual error estimate for the exponent $\nu$.
Those values of $L_{\smin}$ and $\nu$ that minimize $\chi^2/n_{DOF}$ are then used to determine our best estimate of $T_c$.
We note that this method gives a very small statistical error for $T_c$, even negligible in some cases,
but it is important to point out that this error is underestimated when compared to the actual error,
obtained through a true non-linear fit.

%%
%% FIG06: chisquared/ndof
%%

In order to obtain a more realistic (conservative) error bar, we analyze the behavior of both
$\chi^2/n_{DOF}$ and $T_c$ as functions of the fit parameter $1/\nu$, as seen in Fig. \ref{fig:chisquare}.
First we note that all values of $T_c$ in the figure are consistent and the same up to the third decimal place.
We note that an increase of almost $100\%$ in the value of $\chi^2/n_{DOF}$ translates to a fluctuation of almost $4\%$
in the exponent $1/\nu$ which in turn produces a fluctuation of less than $1\%$ in the estimate of $T_c$.
It means that the final value of $T_c$ is not so sensitive to the values of the parameters we fix in the fitting process,
as long as we stay close enough to the value which minimizes $\chi^2/n_{DOF}$.
So, as a criterion to determine the upper and lower bounds of $T_c$,
we considered the values obtained from fits that give $\chi^2/n_{DOF}$ up to $20\%$ larger than the minimum.
It is worth stressing at this point that it is not our goal in this work to obtain a precise description
of the critical behavior for the model.
Therefore, the value of $1/\nu$ is obtained only to achieve a good estimate of $T_c$.

%%%%%%%%%%%%%%%%%%%%%%%%%%%%
%% TAB
\begin{table}[h]
\caption{Estimates of $T_c$ obtained by performing fits
to Eq. (\ref{eq:FSS:tc}) with $L \leq L_\smax$ for different values of $L_\smax$.
As in Fig. \ref{fig:tc} the $T_c(L)$ estimates that serve as input to the fits
correspond to the \emph{maxima} of the layer \textbf{B} magnetic susceptibility $\chi_B$
for $J_{AA}/J_{BB}=0.5$, and $J_{AB}/J_{BB}=-1.0$.}
\label{tab:tc}
\begin{tabular}{ccc}
	\hline\hline
	& \multicolumn{2}{c}{$k_BT_c/J_{BB}$}\\
	\hline
	$L_{\smax}$ & $p=0.6$ & $p=1.0$\\
	\hline
	60  & 1.573(4) & 2.6156(7)\\
	80  & 1.577(2) & 2.6164(5)\\
	110 & 1.578(5) & 2.6163(4)\\
	160 & 1.574(2) & 2.6163(2)\\
	230 & 1.571(2) & 2.6162(1)\\
	320 & 1.572(1) & 2.6161(2)\\
	450 & 1.575(1) & 2.6160(2)\\
	640 & 1.573(1) & 2.61596(7)\\
	900 & 1.575(1) & 2.61593(4)\\
	\hline\hline
\end{tabular}
\end{table}
%%%%%%%%%%%%%%%%%%%%%%%%%%%%

It is intuitive that simulating bigger systems will increase the precision of our results.
Nevertheless, for a fixed amount of computational work, increasing the system size invariably
decreases the amount of simulations we are able to perform.
In test simulations we went as high as $L=900$,
although, since the simulations for the largest lattices are time consuming, we were able to cover only a small
fraction of the parameter space of the system this way.
Since it is our goal to fully explore the parameter space we need to
reach a compromise and keep $L$ as small as possible while still obtaining a fairly accurate result.

The data on Tab. \ref{tab:tc} refer to test simulations performed for $L=10$, 20, 30, 40, 60, 80, 110, 160, 230, 320, 450, 640, and 900.
The table shows estimates of $T_c$ for fits made considering only sizes up to $L_\smax$.
Applying the same $\chi^2/n_{DOF}$ criteria discussed above, we arrive at very consistent estimates of $T_c$
for all values of $L_\smax$. It is clear that the precision increases as we consider bigger sizes;
however the final estimates are the same within error bars.
This means we can have a reasonably good estimate of $T_c$ without so much computational effort.
So, for the determination of $T_c$ in all results presented in the next session, we performed simulations
for $L=10$, 20, 30, 40, 50, 60, 80, and 100.

\section{Results and Discussion}\label{results}

\begin{figure}[h]
\begin{center}
\subfigure[With compensation.\label{fig:magLmax:a}]{
\includegraphics[width=\subfigwidth]{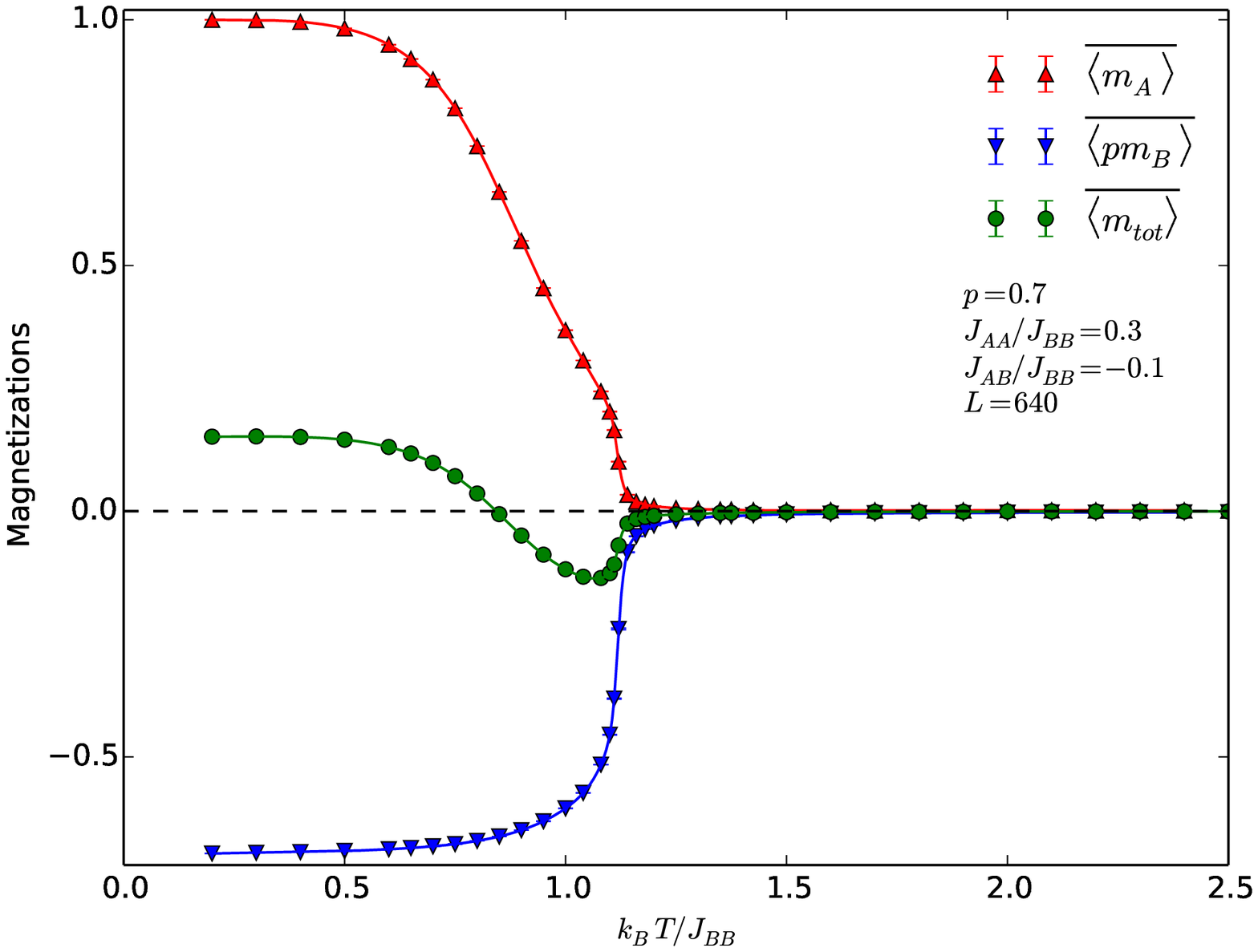}
}
\subfigure[Without compensation.\label{fig:magLmax:b}]{
\includegraphics[width=\subfigwidth]{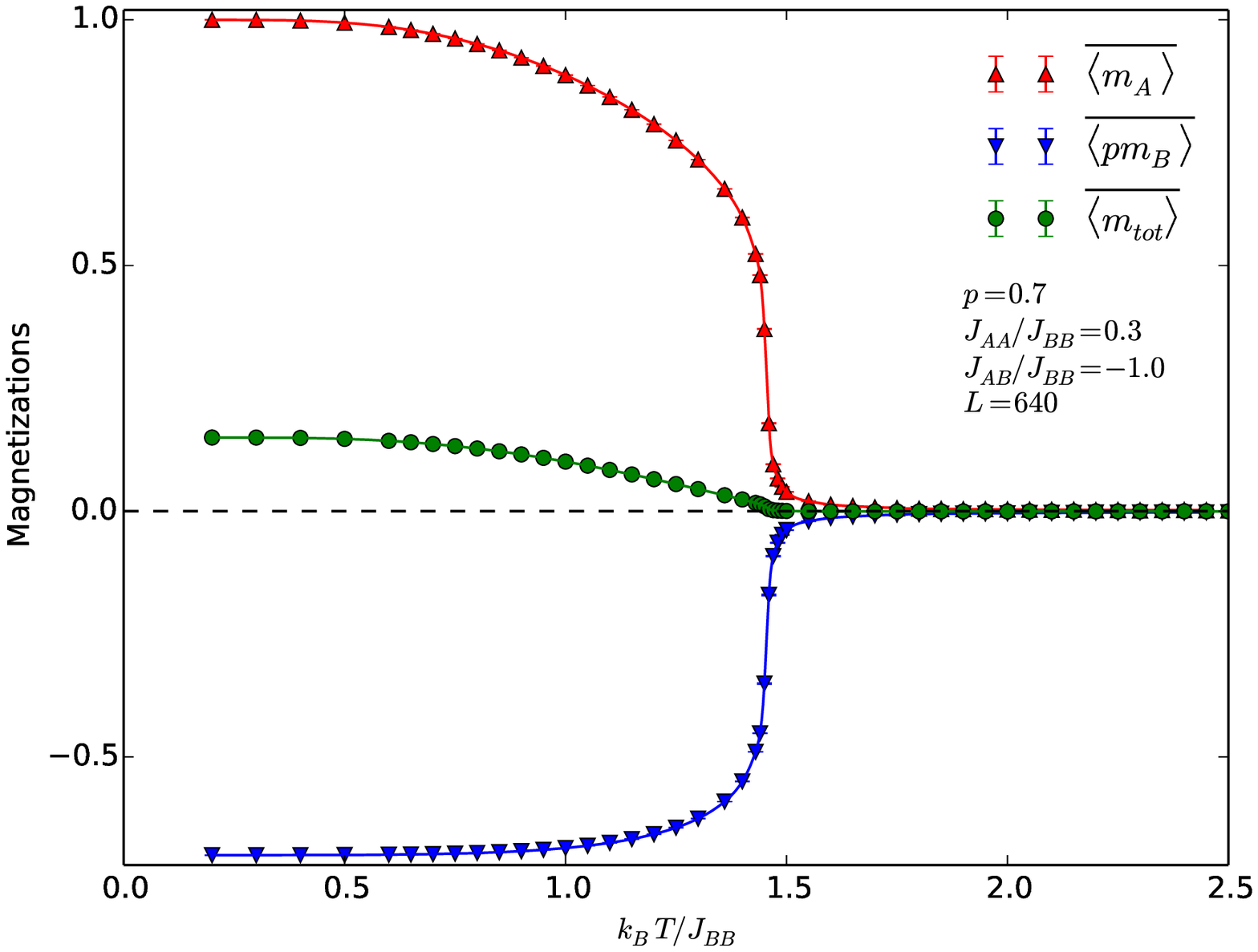}
}
\caption{
\label{fig:magLmax}
Plane magnetizations, $\overline{\mA}$ and $\overline{\pmB}$, and total magnetization $\overline{\m}$
versus the dimensionless temperature $k_BT/J_{BB}$ for $p=0.7$, $J_{AA}/J_{BB}=0.3$ and $L=640$.
Figure (a), for $J_{AB}/J_{BB}=-0.1$, shows a compensation temperature $T_{comp}$
such that $\overline{\m}=0$ and $0<T_{comp}<T_c$
whereas figure (b), for $J_{AB}/J_{BB}=-1.0$, shows no compensation effect.
The symbols correspond to the data and the solid lines are cubic spline interpolations just to guide the eye.
The error bars are smaller than the symbols.
}
\end{center}
\end{figure}

\begin{figure}[h]
\begin{center}
\subfigure[With compensation.\label{fig:mags:a}]{
\includegraphics[width=\subfigwidth]{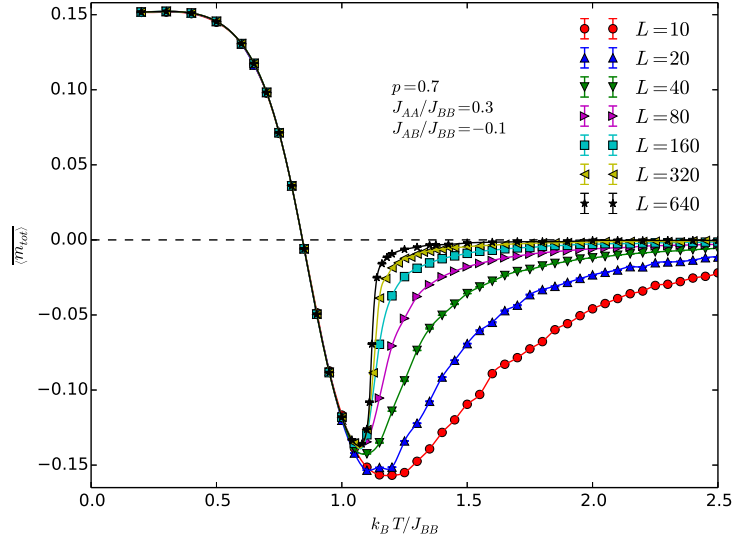}
}
\subfigure[Without compensation.\label{fig:mags:b}]{
\includegraphics[width=\subfigwidth]{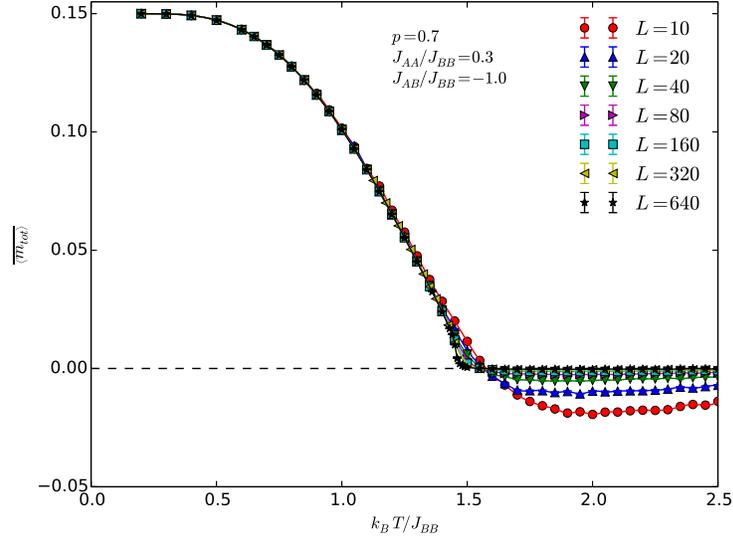}
}
\caption{
\label{fig:mags}
Total magnetization $\overline{\m}$ versus the dimensionless temperature $k_BT/J_{BB}$
for $p=0.7$, $J_{AA}/J_{BB}=0.3$, and several values of system size $L$.
Figure (a), for $J_{AB}/J_{BB}=-0.1$, shows a compensation temperature $T_{comp}$
such that $\overline{\m}=0$ and $0<T_{comp}<T_c$
whereas figure (b), for $J_{AB}/J_{BB}=-1.0$, shows no compensation effect.
The symbols correspond to the data and the solid lines are cubic spline interpolations just to guide the eye.
The error bars are smaller than the symbols.
}
\end{center}
\end{figure}

\begin{figure}[h]
\begin{center}
\includegraphics[width=\figwidth]{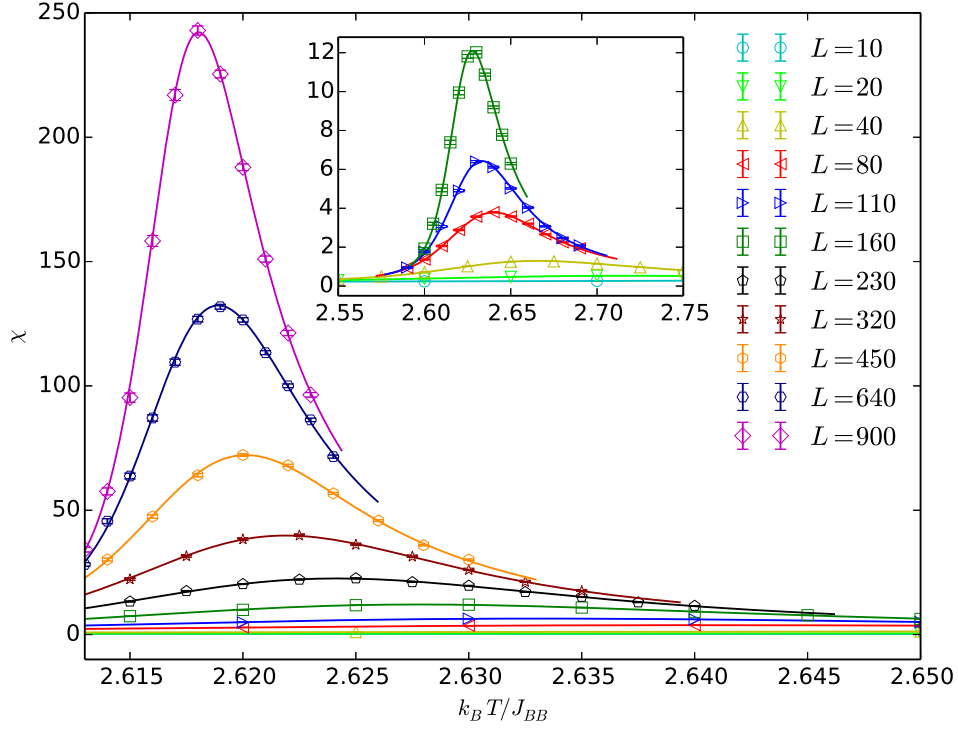}
\caption{
\label{fig:mhist}
Magnetic susceptibility $\chi$ versus the dimensionless temperature $k_BT/J_{BB}$
for $J_{AA}/J_{BB}=0.5$, $J_{AB}/J_{BB}=-1.0$, $p=1.0$, and linear lattice sizes $L$ ranging from 10 to 900.
The inset corresponds to both a zoom in on the vertical axis
and a zoom out on the horizontal axis, in order to better visualize the scaling behavior of the smaller systems.
The symbols correspond to simulation data and the solid lines were obtained using the multiple histogram method.
Where the error bars are not visible, they are smaller than the symbols.
}
\end{center}
\end{figure}

\begin{figure}[h]
\begin{center}
\includegraphics[width=\figwidth]{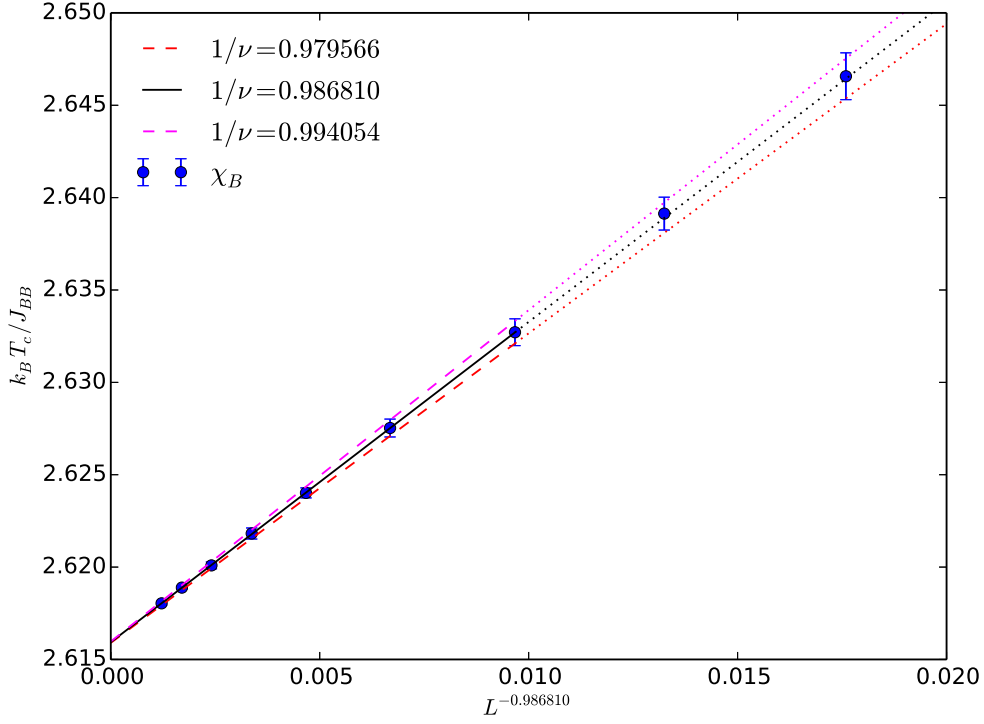}
\caption{
\label{fig:tc}
Dimensionless effective critical temperature $k_BT_c(L)/J_{BB}$ versus $L^{-1/\nu}$
for $J_{AA}/J_{BB}=0.5$, $J_{AB}/J_{BB}=-1.0$, $p=1.0$, and $1/\nu=0.986810$.
The symbols are $T_c(L)$ estimates made by locating the \emph{maxima}
of the layer \textbf{B} magnetic susceptibility $\chi_B$ for $60\leq L\leq 900$.
The solid and dashed lines are fits performed with Eq. (\ref{eq:FSS:tc}) for $L_{\smin}\leq L\leq 900$
and different values of $1/\nu$.
The dotted lines are extrapolations of those fits for $L<L_{\smin}$.
In this particular case, $L_{\smin}=110$ and $1/\nu=0.986810$ are the values which
minimize $\chi^2/n_{DOF}$ and therefore give the best fit.
Where the error bars are not visible, they are smaller than the symbols.
}
\end{center}
\end{figure}

\begin{figure}[h]
\begin{center}
\includegraphics[width=\figwidth]{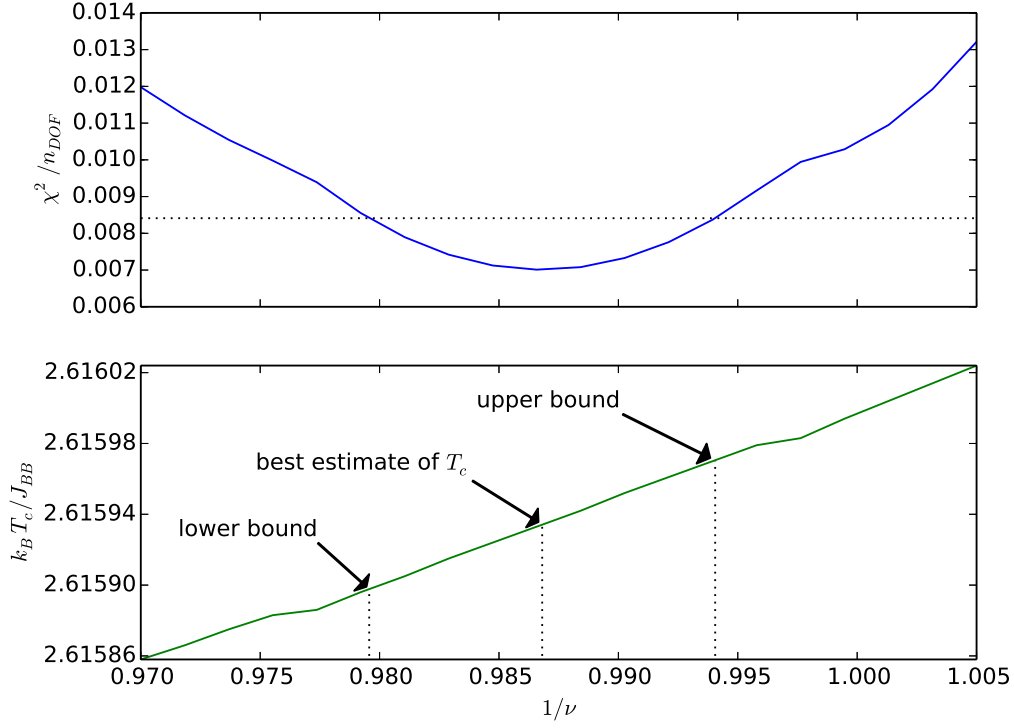}
\caption{
\label{fig:chisquare}
Reduced weighted sum of squared errors $\chi^2/n_{DOF}$ versus $1/\nu$ (above)
and $k_BT_c/J_{BB}$ estimate versus $1/\nu$ (below)
obtained with fits to Eq. (\ref{eq:FSS:tc}).
The $T_c(L)$ estimates that serve as input to the fits correspond to the \emph{maxima}
of the layer \textbf{B} magnetic susceptibility $\chi_B$
for $J_{AA}/J_{BB}=0.5$, $J_{AB}/J_{BB}=-1.0$, $p=1.0$.
The minimum value of $\chi^2/n_{DOF}$ corresponds to the best estimate of $T_c$.
The horizontal dotted line corresponds to the $\chi^2/n_{DOF}$ value
which gives both the lower and upper bounds for $T_c$.
}
\end{center}
\end{figure}

We conducted our study of the magnetic behavior of the bilayer through Monte Carlo simulations.
We have performed simulations and subsequent data analysis
to obtain the critical and compensation temperatures
for different values of the Hamiltonian parameters $J_{AA}/J_{BB}$, $J_{AB}/J_{BB}$, and $p$.
Our goal is to present a detailed account of regions of the parameter space for which the compensation phenomenon is present, 
as seen in Figs. \ref{fig:magLmax:a} and \ref{fig:mags:a}, or absent, as seen in Figs. \ref{fig:magLmax:b} and \ref{fig:mags:b},
and outline the contribution of each parameter for the presence or absence of the aforementioned effect.

Initially, we would like to stress the importance of the asymmetry between the layers of our system.
As far as dilution is concerned, it is trivial to see that there is no compensation effect if $p=0$,
in which case we have a pure two-dimensional ferromagnetic system.
On the other hand, for $p=1$ we have $|\mA|=|\pmB|=1$ at $T=0$.
There are two possibilities:
$(i)$ if $J_{AA}\neq J_{BB}$, we have $|\mA|\neq |\pmB|$ for $0<T<T_c$ and the only temperature
at which the plane magnetizations cancel each other bellow $T_c$ is $T_{comp}=0$;
$(ii)$ if $J_{AA}=J_{BB}$, we have $|\mA|=|\pmB|$ at any $T$ and the system is simply an antiferromagnet.

The roles of intraplanar and interplanar couplings in the compensation phenomenon, however, require additional work and thought.
For instance, Fig. \ref{fig:magLmax} shows the dependence of the planar magnetizations and total magnetization
on the temperature for $L=640$, $p=0.7$ and $J_{AA}/J_{BB}=0.3$.
In Fig. \ref{fig:magLmax:a}, we have $J_{AB}/J_{BB}=-0.1$ and we see a compensation point, such that $\m=0$ and $0<T_{comp}<T_c$,
whereas Fig. \ref{fig:magLmax:b}, for $J_{AB}/J_{BB}=-1.0$, shows no compensation effect.
This indicates that the occurrence of the compensation temperature is favoured
by weaker interplanar couplings, as in this case the only difference between a system with compensation
(Fig. \ref{fig:magLmax:a}) and without compensation (Fig. \ref{fig:magLmax:b}) is the value of $J_{AB}/J_{BB}$.
It is easily seen that, if $|J_{AB}/J_{BB}|\gg 1$, first-neighbor spins on different planes will be ``frozen''
at different states (one $+1$ and the other $-1$) and, since $p<1$, $|\mA|>|\pmB|$ for $T<T_c$.
Therefore a strong $J_{AB}$ coupling rules out the presence of the compensation effect.

It is also easy to realize that the diluted lattice needs to have a stronger intraplanar coupling for the compensation effect to occur.
Consider an extreme case in which $J_{AB}/J_{BB}=0$, i. e., we have two independent ferromagnetic systems,
each undergoing a phase transition at a different critical temperature,
let us say $T_{c,A}$ and $T_{c,B}$ for lattices \textbf{A} and \textbf{B} respectively.
Consider also that we still have a way other than the antiferromagnetic interplanar coupling to keep the planar magnetizations opposed to one another.
If both lattices have no dilution and $J_{AA}>J_{BB}$, we have $T_{c,A}>T_{c,B}$,
so $|\mA|=|\pmB|=1$ at $T=0$ and $|\mA|>|\pmB|$ for $0>T>T_{c,A}$,
i. e., there is no temperature other than zero where $|\mA|$ and $|\pmB|$ are the same.
The effect of dilution on lattice \textbf{B} is to lower the $|\pmB|$ curve such that $|\pmB|<|\mA|$ at $T=0$
while simultaneously lowering $T_{c,B}$, so we still have $|\mA|>|\pmB|$ for $0<T<T_{c,A}$
and the $|\mA|$ and $|\pmB|$ curves do not cross at any temperature, even $T=0$.

For the interacting system we can draw a similar conclusion.
The presence of an intraplanar coupling, no matter how weak, forces both lattices to behave as a single system and have a unique transition temperature, $T_c$,
which is greater than both $T_{c,A}$ and $T_{c,B}$ of the non-interacting systems mentioned above.
If both lattices have no dilution, we have $|\mA|=|\pmB|=1$ at $T=0$ and, as discussed above,
the system will either be a ferrimagnet with $T_{comp}=0$ (if $J_{AA}\neq J_{BB}$) or an antiferromagnet (if $J_{AA}=J_{BB}$).
As in the case of the non-interacting system, the effect of dilution is still to lower the $|\pmB|$ curve such that $|\pmB|<|\mA|$ at $T=0$.
If $J_{AA}>J_{BB}$, we have $|\mA|>|\pmB|$ for $0\leq T<T_c$, therefore in this case we have no compensation temperature for any $p>0$.
Even for $J_{AA}<J_{BB}$, the compensation effect will only be present if
both $J_{AA}/J_{BB}$ and $|J_{AB}/J_{BB}|$ remain small enough that the $|\mA|$ curve drops more gradually toward zero than $|\pmB|$,
as can be seen in Fig. \ref{fig:magLmax:a}.

\begin{figure}[h]
\begin{center}
\includegraphics[width=\figwidth]{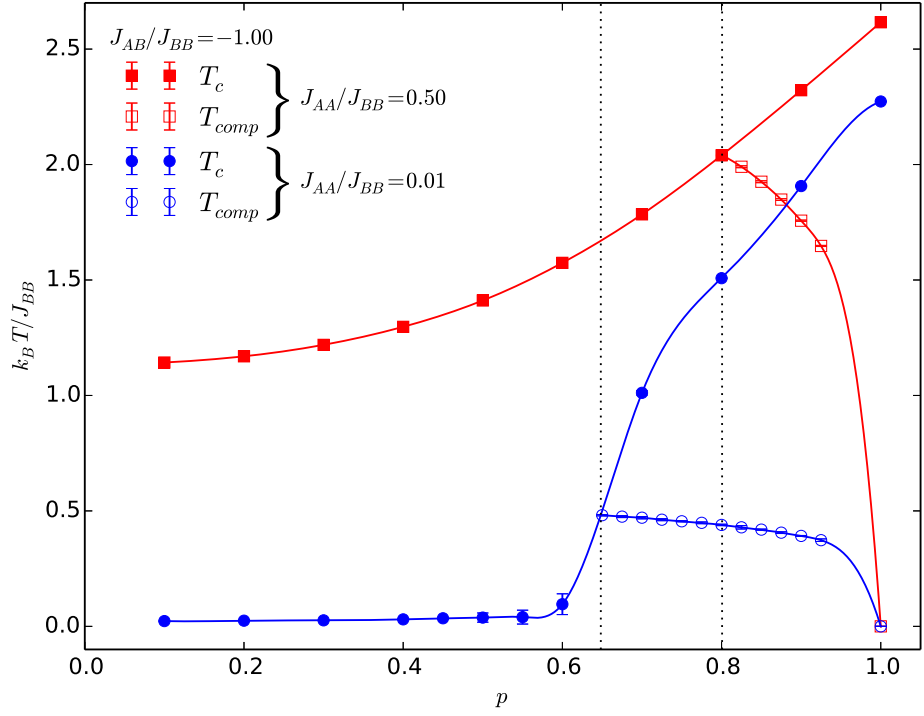}
\caption{
\label{fig:07}
Critical temperatures $T_c$ (filled symbols) and compensation temperatures
$T_{comp}$ (empty symbols) versus concentration $p$.
The solid lines are cubic spline interpolations just to guide the eye.
The vertical dotted lines mark the characteristic concentration $p^\star$ where $T_c=T_{comp}$.
Where the error bars are not visible, they are smaller than the symbols.
}
\end{center}
\end{figure}

In Fig. \ref{fig:07},
we plot the critical temperatures and compensation temperatures as functions of the concentration $p$.
The solid symbols are the critical temperatures and the empty ones are the compensation temperatures.
To draw the solid lines we use cubic spline interpolations just as a guide to the eye.
The vertical dotted lines mark the characteristic concentration $p^\star$ for which the $T_c$ and $T_{comp}$ curves meet.
For $p<p^\star$, for a given set of parameter values, the compensation phenomenon does not occur.

\begin{figure}[h]
\begin{center}
\includegraphics[width=\figwidth]{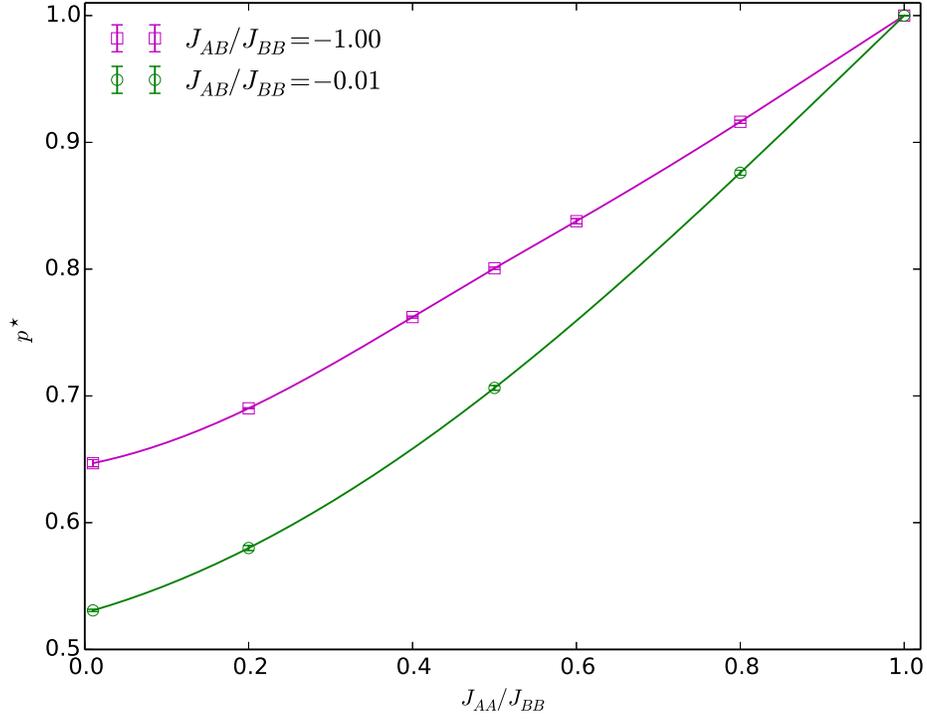}
\caption{
\label{fig:08}
Characteristic concentration $p^\star$, bellow which there is no compensation,
versus the ratio $J_{AA}/J_{BB}$ for $J_{AB}/J_{BB}=-0.01$ (circles) and $J_{AB}/J_{BB}=-1.0$ (squares).
The solid lines are cubic spline interpolations just to guide the eye.
Where the error bars are not visible, they are smaller than the symbols.
}
\end{center}
\end{figure}

\begin{figure}[h]
\begin{center}
\includegraphics[width=\figwidth]{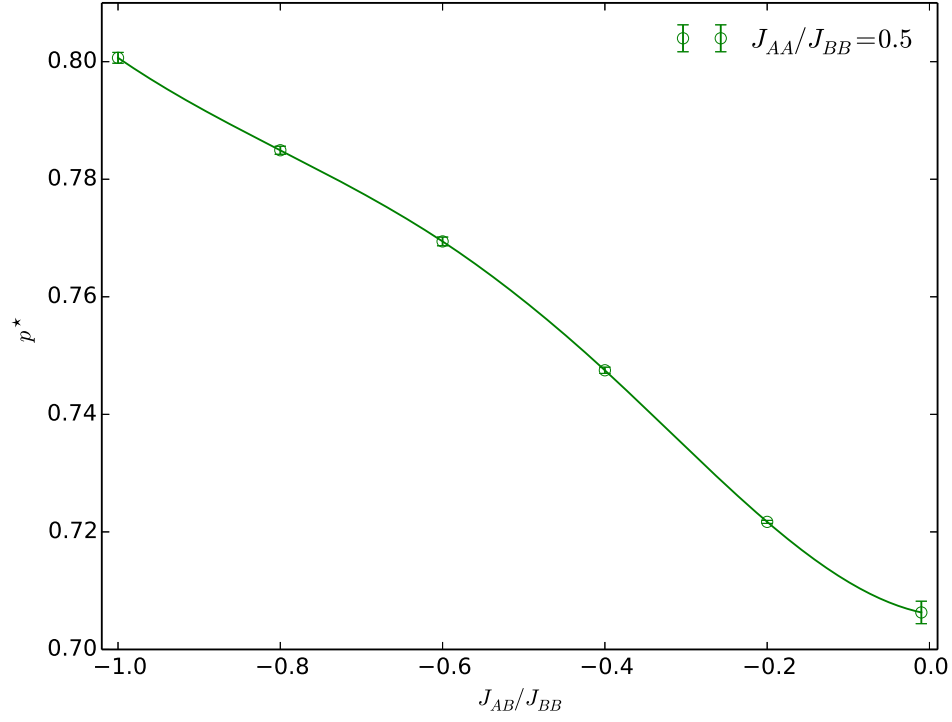}
\caption{
\label{fig:09}
Characteristic concentration $p^\star$, bellow which there is no compensation,
versus the ratio $J_{AB}/J_{BB}$ for $J_{AA}/J_{BB}=0.5$.
The solid lines are cubic spline interpolations just to guide the eye.
Where the error bars are not visible, they are smaller than the symbols.
}
\end{center}
\end{figure}

We also notice in Fig. \ref{fig:07} that $p^\star$ is higher for $J_{AA}/J_{BB}=0.5$ than it is for $J_{AA}/J_{BB}=0.01$,
indicating that the concentration $p^\star$ increases as the interaction within layer \textbf{A} becomes stronger.
This tendency is confirmed in Fig. \ref{fig:08},
where we plot the characteristic concentration $p^\star$ as a function of the ratio $J_{AA}/J_{BB}$
for a weak interplanar coupling ($J_{AB}/J_{BB}=-0.01$) and a strong one ($J_{AB}/J_{BB}=-1.0$),
and in both cases $p^\star$ increases with a stronger lattice \textbf{A} intraplanar coupling.
Fig. \ref{fig:09} also reveals a similar tendency for the behavior of $p^\star$ as a function of the interplanar coupling:
we see that the concentration $p^\star$ decreases as $J_{AB}/J_{BB}$ increases
(but it should be noted that as the interplanar interaction is antiferromagnetic,
it also means that $p^\star$ increases as the coupling gets stronger).

\begin{figure}[h]
\begin{center}
\includegraphics[width=\figwidth]{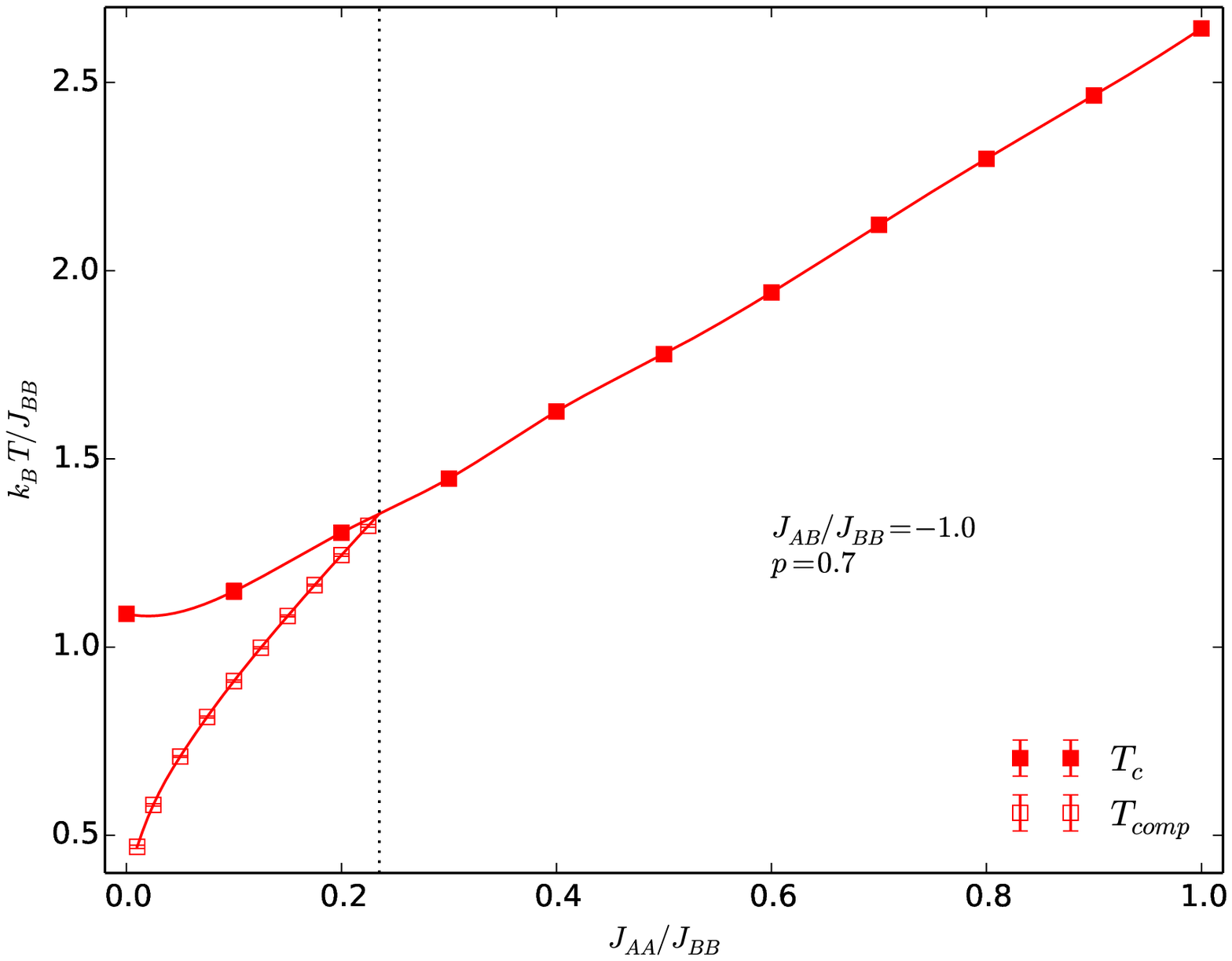}
\caption{
\label{fig:10}
Critical temperatures $T_c$ (filled symbols) and compensation temperatures $T_{comp}$ (empty symbols)
versus the ratio $J_{AA}/J_{BB}$ for $p=0.7$ and $J_{AB}/J_{BB}=-1.0$.
The solid lines are cubic spline interpolations just to guide the eye.
The vertical dotted line mark the characteristic ratio $(J_{AA}/J_{BB})^\star$ where $T_c=T_{comp}$.
Where the error bars are not visible, they are smaller than the symbols.
}
\end{center}
\end{figure}

\begin{figure}[h]
\begin{center}
\includegraphics[width=\figwidth]{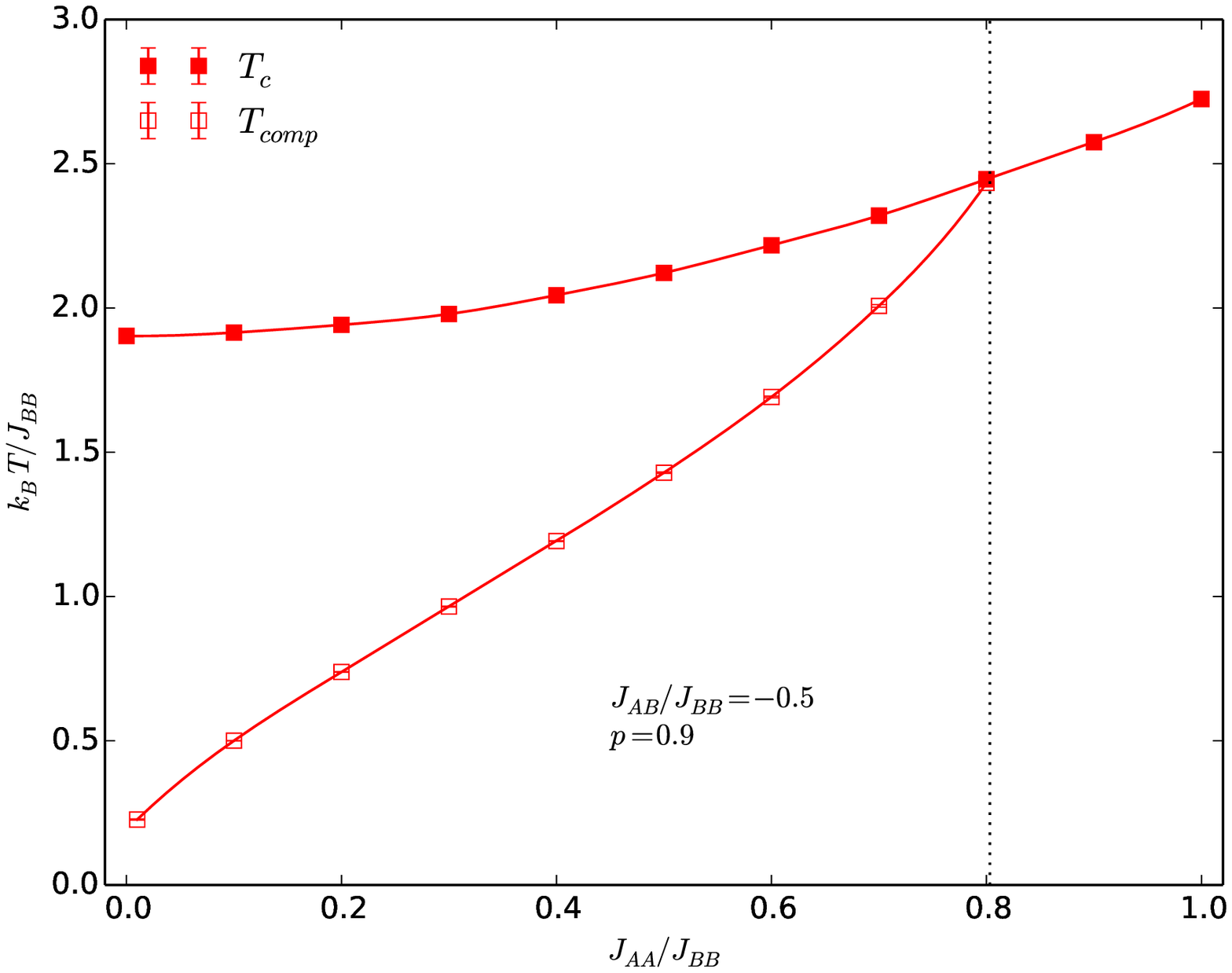}
\caption{
\label{fig:11}
Critical temperatures $T_c$ (filled symbols) and compensation temperatures $T_{comp}$ (empty symbols)
versus the ratio $J_{AA}/J_{BB}$ for $p=0.9$ and $J_{AB}/J_{BB}=-0.5$.
The solid lines are cubic spline interpolations just to guide the eye.
The vertical dotted line mark the characteristic ratio $(J_{AA}/J_{BB})^\star$ where $T_c=T_{comp}$.
Where the error bars are not visible, they are smaller than the symbols.
}
\end{center}
\end{figure}

\begin{figure}[h]
\begin{center}
\includegraphics[width=\figwidth]{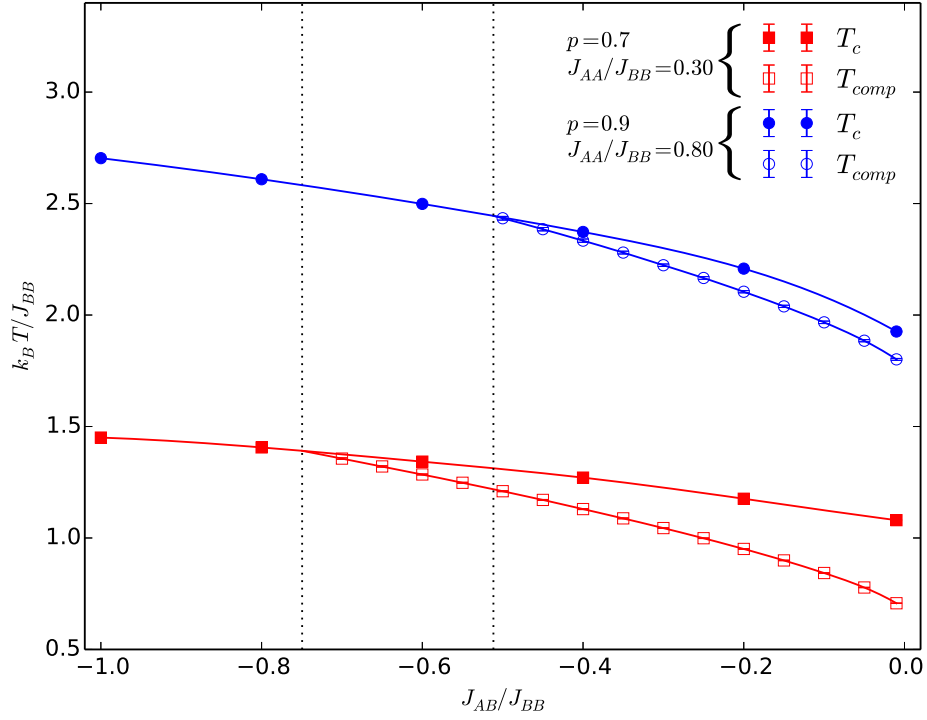}
\caption{
\label{fig:12}
Critical temperatures $T_c$ (filled symbols) and compensation temperatures $T_{comp}$ (empty symbols)
versus the ratio $J_{AB}/J_{BB}$ for $p=0.7$ and $J_{AA}/J_{BB}=0.3$ (squares) and $p=0.9$ and $J_{AA}/J_{BB}=0.8$ (circles).
The solid lines are cubic spline interpolations just to guide the eye.
The vertical dotted lines mark the characteristic ratio $(J_{AB}/J_{BB})^\star$ where $T_c=T_{comp}$.
Where the error bars are not visible, they are smaller than the symbols.
}
\end{center}
\end{figure}

For a fixed value of $p$, we analyze how the parameters $J_{AA}/J_{BB}$ and $J_{AB}/J_{BB}$
influence the absence or presence of the compensation phenomenon.
In Figs. \ref{fig:10} and \ref{fig:11},
we plot the critical temperature, $T_c$ (filled symbols), and the compensation temperature, $T_{comp}$ (empty symbols),
as functions of $J_{AA}/J_{BB}$ for fixed values of $p$ and $J_{AB}/J_{BB}$.
In both cases, the vertical dotted line marks the characteristic ratio $(J_{AA}/J_{BB})^\star$ where the $T_c$ and $T_{comp}$ curves meet
and, above which, there is no compensation.
In Fig. \ref{fig:10} we see that for $p=0.7$ and a strong interplanar coupling
we only have compensation for the lower values of $J_{AA}/J_{BB}$.
However, in Fig. \ref{fig:11} the increase in active site concentration from $p=0.7$ to $0.9$
widens the range of $J_{AA}/J_{BB}$ for which we have compensation,
even though the interplanar interaction strength's absolute value has been reduced from
$|J_{AB}/J_{BB}|=1.0$ to $|J_{AB}/J_{BB}|=0.5$.
In Fig. \ref{fig:12} we explore the 
dependence of the critical temperature, $T_c$ (filled symbols), and the compensation temperature, $T_{comp}$ (empty symbols),
on the ratio $J_{AB}/J_{BB}$ for fixed values of $p$ and $J_{AA}/J_{BB}$.
Again, the vertical dotted lines mark the characteristic ratio $(J_{AB}/J_{BB})^\star$ where the $T_c$ and $T_{comp}$ curves meet.
It is interesting to point out that a weaker interplanar coupling favors the compensation effect although,
at the same time, it reduces the critical temperature.

\begin{figure}[h]
\begin{center}
\subfigure[$p=0.7$.\label{fig:13}]{
\includegraphics[width=\subfigwidth]{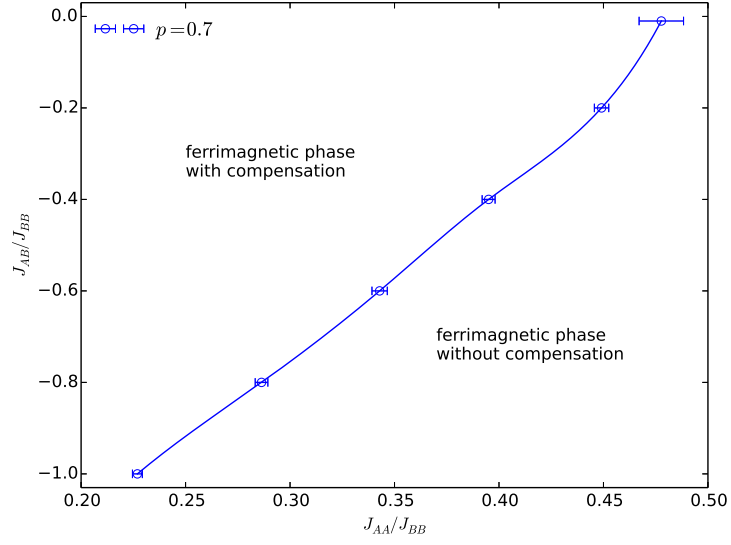}
}
\subfigure[$p=0.9$.\label{fig:14}]{
\includegraphics[width=\subfigwidth]{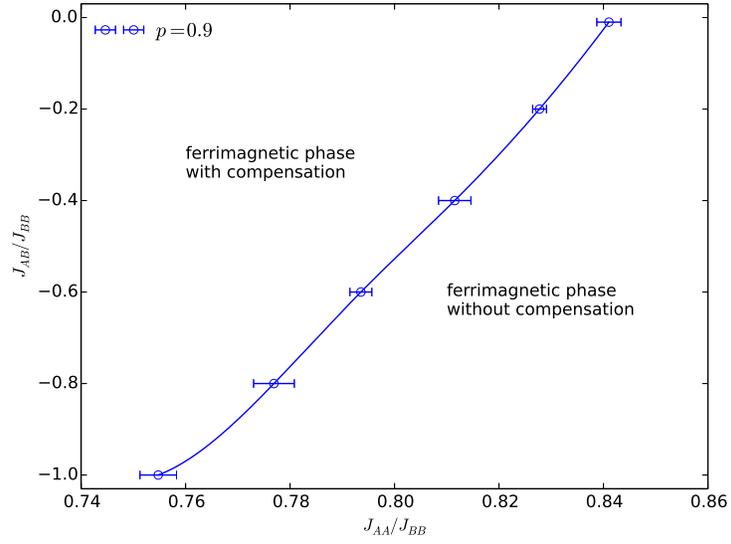}
}
\caption{
\label{fig:13-14}
Phase diagrams.
The solid line is a cubic spline interpolation just to guide the eye.
}
\end{center}
\end{figure}

As it follows from the results presented above,
we can divide the parameter space of our Hamiltonian in two distinct areas of interest.
One is a ferrimagnetic phase for which there is no compensation at any temperature,
while the other corresponds to a ferrimagnetic phase for which the compensation phenomenon takes place at a certain temperature $T_{comp}$.
Therefore, it may be useful to present some phase diagrams showing the areas corresponding to the existence of both phases.
To this end, Fig. \ref{fig:13} shows the $J_{AB}/J_{BB}$ versus $J_{AA}/J_{BB}$ phase diagram, as obtained from our numerical results.
Above the depicted line the system presents a ferrimagnetic phase with compensation temperature,
while below the line no compensation effect is present, although the system is also in an ordered state.
This diagram confirms that the compensation effect is favored by a weaker interplanar coupling and by a more
pronounced intraplanar coupling asymmetry.
An analogous phase diagram for $p=0.9$ is presented in Fig. \ref{fig:14}.
It is evident that, as we reduce the atomic dilution in layer \textbf{B},
the critical value of interplanar coupling depends less and less on the value of the intraplanar interactions in layer \textbf{A}.
Moreover, as $p$ increases, the range occupied by the ferrimagnetic phase without compensation greatly decreases.

Our MC calculations can also be compared to the mean-field-like pair approximation method (PA)
presented in Ref. \onlinecite{balcerzak2014ferrimagnetism}.
Although our results agree qualitatively with the PA,
as it is evident if we compare, for example,
Figs. \ref{fig:07}, \ref{fig:10}, and \ref{fig:11} in the present paper
with their counterparts in Ref. \onlinecite{balcerzak2014ferrimagnetism},
Figs. 2, 6, and 7, respectively,
our quantitative results differ quite drastically from the PA.

It is necessary to bear in mind that, in order to compare the values of $T_c$ and $T_{comp}$
in this work with those in Ref. \onlinecite{balcerzak2014ferrimagnetism},
we need to acknowledge the fact that in the later the Ising spin
variables assume the values $s_i=\pm1/2$, whereas in our simulations we used $s_i=\pm 1$.
This alone is responsible for a difference by a factor of four
between the energy scales, and consequently between the temperature scales, in the two works,
i. e., all temperatures presented here have to be divided by four to be compared with
the temperatures presented in Ref. \onlinecite{balcerzak2014ferrimagnetism}.
Following this simple procedure, the comparison between our Fig. \ref{fig:07}
and its PA counterpart (Fig. 2 in Ref. \onlinecite{balcerzak2014ferrimagnetism}),
for $J_{AA}/J_{BB}=0.5$ and $J_{AB}/J_{BB}=-1.0$,
leads to $k_BT_c/J_{BB}=2.040(1)$ at $p=0.8$ for the MC estimate
and $k_BT_c/J_{BB}\approx 0.65$ for the PA estimate at the same concentration.
Correcting for the difference in temperature scale,
it follows that the PA critical temperature is
approximately $27\%$ higher than our MC $T_c$ estimate for this particular choice of parameters.
The same logic applies to other estimates of both $T_c$ and $T_{comp}$
and we consistently find that PA gives slightly higher estimates than MC,
as it is expected from a mean-field-like method \cite{livro:julia}.

Still concerning Fig. \ref{fig:07}, it is worth pointing out that our figures for $p^\star$ are different from the PA estimates.
Namely, the MC estimate is slightly lower than the PA figure for $J_{AA}/J_{BB}=0.01$ and slightly higher for $J_{AA}/J_{BB}=0.5$.
Therefore, the effect of changing $J_{AA}/J_{BB}$ is stronger in MC simulations, as regards the value of $p^\star$.
Similarly, in Fig. \ref{fig:10} we obtain a $(J_{AA}/J_{BB})^\star$
approximately $35\%$ smaller than the value we obtain from Fig. 6 in Ref. \onlinecite{balcerzak2014ferrimagnetism}
whereas Fig. \ref{fig:11} gives a value of $(J_{AA}/J_{BB})^\star$
slightly lower than the one from Fig. 7 in Ref. \onlinecite{balcerzak2014ferrimagnetism}.
%%
%% our Fig. 10 JAA/JBB=0.233 ... their Fig. 6 JAA/JBB = 0.36 :: dif = -35.277 percent
%%

Fig. 8 in Ref. \onlinecite{balcerzak2014ferrimagnetism}
further helps to highlight the differences between PA and MC results as for both sets of parameters,
$(J_{AA}/J_{BB}=0.2; p=0.6)$ and $(J_{AA}/J_{BB}=0.5; p=0.7)$,
PA results show the existence of a compensation temperature above a particular value of the ratio $J_{AB}/J_{BB}$
whereas in our MC calculations, for $p=0.7$, we found no compensation for any $J_{AB}/J_{BB}$ whatsoever (see Fig. \ref{fig:13}).
In order to present an analogous diagram, with the presence of a phase with compensation,
in this work we had to chose completely different sets of parameters, as seen in our Fig. \ref{fig:12}.
The same phenomenon happens with Fig. 9 in Ref. \onlinecite{balcerzak2014ferrimagnetism},
which clearly shows the occurrence of the compensation phenomenon for
the parameters $p=0.6$, $J_{AA}/J_{BB}=0.2$, and $J_{AB}/J_{BB}=-0.1$ for the PA
while there is no compensation for the same parameter set in MC.
So, for the analogous of Figs. 9 and 10 in Ref. \onlinecite{balcerzak2014ferrimagnetism},
we also chose a different set of values for the parameters,
as seen in our Figs. \ref{fig:magLmax} and \ref{fig:mags}.

Finally, the phase diagrams presented in Figs. \ref{fig:13} and \ref{fig:14}
may also be compared to Figs. 11 and 12 in Ref. \onlinecite{balcerzak2014ferrimagnetism}, respectively.
In both cases we see that, for fixed values of $p$ and $J_{AB}/J_{BB}$,
the $J_{AA}/J_{BB}$ values that fall in the line separating the phases with and without compensation
are lower in Monte Carlo than in the mean-field-like approximation.
Also, in both cases the difference grows as $|J_{AB}/J_{BB}|$ increases.
For $p=0.7$, at $J_{AB}/J_{BB}=-0.01$, the MC value for $J_{AA}/J_{BB}$
is approximately $17\%$ lower than the PA estimate while at $J_{AB}/J_{BB}=-1.0$
the MC result gets approximately $41\%$ lower.
On the other hand, for $p=0.9$, the discrepancy ranges from only $\approx 1.5\%$
at $J_{AB}/J_{BB}=-0.01$ to a still small $\approx 6.9\%$ at $J_{AB}/J_{BB}=-1.0$,
which indicates that the discrepancy is more pronounced for lower concentrations (or higher dilutions).

%at $J_{AB}/J_{BB}=-1.0$, the MC value for $(J_{AA}/J_{BB})^\star$ is $\approx 17\%$ lower than the PA estimate
%while at $J_{AB}/J_{BB}=-0.01$ the MC resulta is $\approx 41\%$ lower.
%For it ranges from $\approx 1.5\%$ to  $\approx 6.9\%$ difference
%%p=0.7
%%-41.02564102564102; JAB=-1.0 
%%-17.272727272727277; JAB=-0.01
%%p=0.9
%%-6.905055487053026; JAB=-1.0
%%-1.524032825322393; JAB=-0.01

\section{Conclusion}\label{conclusion}

In this work we studied a bilayer ferrimagnetic model consisting of two
interacting square lattices. The intralayer couplings are ferromagnetic while 
the interlayer interactions are antiferromagnetic. One of the layers is diluted, with
sites randomly chosen to be magnetic or non-magnetic with probability $p$ or $(1-p)$,
respectively. Our main goal is to obtain the conditions for the presence of compensation
temperatures, $T_{comp}$. In these temperatures, the magnetization of the two layers
cancel out below the critical temperature $T_c$.

Previous studies on this model used a mean-field-like approximation, which may not
correctly represent two-dimensional models or may fail in predicting accurate values
for some physical quantities. Additionally, a precise evaluation of the regions where
the compensation effect takes place is an invaluable information for experimentalists.
With that in mind, we employed Monte Carlo simulations, using Metropolis and Wolff
algorithms, accompanied by the multiple histogram reweighting method and finite-size scaling tools.
Applied to lattices of different linear sizes $L$,
these approaches allow for a very precise evaluation of both $T_{comp}$ and $T_c$.

Since our objective in this work is not to obtain an accurate determination of
critical exponents, it was not necessary to resort to large lattices, specially when
$T_{comp}$ was not close to $T_c$. Studying lattices with linear sizes $L \leq 900$ we
determined the range in the Hamiltonian parameters for which the compensation effect
is present. The critical temperature is obtained from the scaling relation shown
in Eq. (\ref{eq:FSS:tc}), combined with a procedure describe in Refs. \onlinecite{artigo:landau}
and \onlinecite{diaz2012feru}.

Initially, we provide general arguments to support the necessity of both dilution
and a stronger intralayer interaction within the diluted layer (when compared to the 
interaction within the other layer), in order that the compensation effect may be possible.
We then study the effects of other parameters using two categories
of phase diagrams: $(i)$ the dependence of $T_c$ and $T_{comp}$ on the Hamiltonian
parameters (namely, $p$, $J_{AA}$, and $J_{AB}$) and $(ii)$ the dependence of $p^\star$ 
(the value of $p$ below which there is no compensation effect) on $J_{AA}$ and $J_{AB}$. 
In  $(ii)$ it is shown that the region with compensation effect diminishes with either
increasing $J_{AA}$ or $|J_{AB}|$, while in $(i)$ we can see that both $T_c$ and $T_{comp}$
increase with $J_{AA}$ or $|J_{AB}|$, such that, eventually, $T_{comp}=T_c$ and, for
greater values of $J_{AA}$ and $|J_{AB}|$, the compensation effect is not present anymore.

A summary of the results is then depicted in a convenient way on 
$J_{AB} \times J_{AA}$ diagrams, which confirm that the compensation effect is favored 
by weaker interplanar couplings and by a more pronounced intraplanar coupling asymmetry. 
It is worth noting that these diagrams show behaviors considerably different 
from the ones observed in the mean-field-like approximation applied to the same model \cite{balcerzak2014ferrimagnetism}.

Work is now underway to generalize the present model to study spins with
continuous symmetry (the Heisenberg model), multilayers, and the evaluation of
critical exponents for the latter (where dilution is a relevant parameter, in the
renormalization-group sense).

%In this study we ...
%\begin{itemize}
%\item Important:::
%\begin{itemize}
%\item Our results are fucking awsome
%\item We should not rely on mean-field results to predict any experimental outcome
%\item Instead, we NEED the Monte Carlo results for that
%\end{itemize}
%\item Perspectives:::
%\begin{itemize}
%\item Fazer o mesmo que foi feito nesse trabalho agora para o sistema 3D (ja em andamento)
%\item Critical expoents
%\item Critical expoents of the 3D system (multilayer)
%\item Heisenberg spins
%\end{itemize}
%\end{itemize}

\begin{acknowledgments}
This work has been partially supported by Brazilian Agencies FAPESC, CNPq, and CAPES.
\end{acknowledgments}

\bibliography{biblio}

%%%%%%%%%%
%%% FIGS.
%%%%%%%%%%
\newpage

%%%%%%%%%%%%%%%%%%%%%%%%%%%%%%%
%%% FIG 1
%%%%%%%%%%%%%%%%%%%%%%%%%%%%%%%
%%%%%%%%%%%%%%%%%%%%%%%%%%%%%%%%

%%%%%%%%%%%%%%%%%%%%%%%%%%%%%%%
%%% FIG 2
%%%%%%%%%%%%%%%%%%%%%%%%%%%%%%%
%%%%%%%%%%%%%%%%%%%%%%%%%%%%%%%%

%%%%%%%%%%%%%%%%%%%%%%%%%%%%%%%
%%% FIG 3
%%%%%%%%%%%%%%%%%%%%%%%%%%%%%%%
%%%%%%%%%%%%%%%%%%%%%%%%%%%%%%%

%%%%%%%%%%%%%%%%%%%%%%%%%%%%%%%
%%% FIG 4
%%%%%%%%%%%%%%%%%%%%%%%%%%%%%%%
%%%%%%%%%%%%%%%%%%%%%%%%%%%%%%%

%%%%%%%%%%%%%%%%%%%%%%%%%%%%%%%
%%% FIG 5
%%%%%%%%%%%%%%%%%%%%%%%%%%%%%%%
%%%%%%%%%%%%%%%%%%%%%%%%%%%%%%%

%%%%%%%%%%%%%%%%%%%%%%%%%%%%%%%
%%% FIG 6
%%%%%%%%%%%%%%%%%%%%%%%%%%%%%%%
%%%%%%%%%%%%%%%%%%%%%%%%%%%%%%%

%%%%%%%%%%%%%%%%%%%%%%%%%%%%%%%
%%% FIG 7
%%%%%%%%%%%%%%%%%%%%%%%%%%%%%%%
%%%%%%%%%%%%%%%%%%%%%%%%%%%%%%%

%%%%%%%%%%%%%%%%%%%%%%%%%%%%%%%
%%% FIG 8
%%%%%%%%%%%%%%%%%%%%%%%%%%%%%%%
%%%%%%%%%%%%%%%%%%%%%%%%%%%%%%%

%%%%%%%%%%%%%%%%%%%%%%%%%%%%%%%
%%% FIG 09
%%%%%%%%%%%%%%%%%%%%%%%%%%%%%%%
%%%%%%%%%%%%%%%%%%%%%%%%%%%%%%%

%%%%%%%%%%%%%%%%%%%%%%%%%%%%%%%
%%% FIG 10
%%%%%%%%%%%%%%%%%%%%%%%%%%%%%%%
%%%%%%%%%%%%%%%%%%%%%%%%%%%%%%%

%%%%%%%%%%%%%%%%%%%%%%%%%%%%%%%
%%% FIG 11
%%%%%%%%%%%%%%%%%%%%%%%%%%%%%%%
%%%%%%%%%%%%%%%%%%%%%%%%%%%%%%%

%%%%%%%%%%%%%%%%%%%%%%%%%%%%%%%
%%% FIG 12
%%%%%%%%%%%%%%%%%%%%%%%%%%%%%%%
%%%%%%%%%%%%%%%%%%%%%%%%%%%%%%%

%%%%%%%%%%%%%%%%%%%%%%%%%%%%%%%
%%% FIGS 13 (a) and (b)
%%%%%%%%%%%%%%%%%%%%%%%%%%%%%%%
%%%%%%%%%%%%%%%%%%%%%%%%%%%%%%%

%%%%%%%%%%%%%%%%%%%%%%%%%%%%%%%
%%% FIG 13
%%%%%%%%%%%%%%%%%%%%%%%%%%%%%%%
%\begin{figure}[h]
%\begin{center}
%\includegraphics[width=\figwidth]{fig13}
%\caption{
%\label{fig:13}
%The phase diagram for concentration $p=0.7$.
%The solid line is a cubic spline interpolation just to guide the eye.
%}
%\end{center}
%\end{figure}
%%%%%%%%%%%%%%%%%%%%%%%%%%%%%%%
%%%%%%%%%%%%%%%%%%%%%%%%%%%%%%%
%%% FIG 14
%%%%%%%%%%%%%%%%%%%%%%%%%%%%%%%
%\begin{figure}[h]
%\begin{center}
%\includegraphics[width=\figwidth]{fig14}
%\caption{
%\label{fig:14}
%The phase diagram for concentration $p=0.9$.
%The solid line is a cubic spline interpolation just to guide the eye.
%}
%\end{center}
%\end{figure}
%%%%%%%%%%%%%%%%%%%%%%%%%%%%%%%

\end{document}